\documentclass[12pt]{iopart}
\usepackage{graphicx}      

\begin{document}
\title{Numerical approach to the dynamical Casimir effect}
\author{Marcus Ruser} 

\address{D{\'e}partement de Physique Th{\'e}orique, Universit{\'e}
de Gen{\`e}ve, 24 quai Ernest Ansermet, CH-1211 Gen{\`e}ve 4, Switzerland}
\ead{Marcus.Ruser@physics.unige.ch}

\begin{abstract}
The dynamical Casimir effect for a massless scalar field in 1+1-dimensions is studied 
numerically by solving a system of coupled first-order differential equations. 
The number of scalar particles created from vacuum is given by the solutions to 
this system which can be found by means of standard numerics. The formalism 
already used in a former work is derived in detail and is applied to resonant 
as well as off-resonant cavity oscillations.   
\end{abstract}


\section{Introduction}
The possibility of creating photons out of vacuum fluctuations of 
the quantized electromagnetic field in dynamical cavities, 
the so-called dynamical Casimir effect (see \cite{Dodonov:2001a}
for a review), demonstrates the highly non-trivial nature of
the quantum vacuum. 

A scenario of particular interest are so-called vibrating cavities 
\cite{Lambrecht:1996} where the distance between two parallel (ideal) 
mirrors changes periodically in time. The occurrence of resonance effects 
between the mechanical motion of the mirror and the quantum vacuum
leading to (even exponentially) increasing occupation numbers  
in the resonance modes makes this configuration the most 
promising candidate for an experimental verification of this pure quantum effect.

Particle creation in one-dimensional vibrating cavities has been studied in 
numerous works \cite{Dodonov:1993,Dodonov:1996,Dodonov:1996a,Ji:1997,
Schuetzhold:1998, Dodonov:1998, Klimov:1997,Fu:1997,
Chizhov:1997}. When considering small amplitude oscillations,
analytical results can be deduced showing that under 
resonance conditions the particle occupation numbers 
increase quadratically in time. Particle creation due 
to off-resonant wall motions has been investigated in, 
e.g., \cite{Dodonov:1998}. The evolution of the energy density 
in a one-dimensional cavity with one vibrating wall has also been 
studied by many authors \cite{Law:1994,Cole:1995,
Meplan:1996,Dalvit:1998,Rog:2001,Andreata:2000,Llave:1999} demonstrating
that the total energy inside a resonantly vibrating cavity grows 
exponentially in time (see also \cite{Dodonov:1996}) while the total 
particle number increases only quadratically. Thus a pumping of energy 
into higher frequency modes takes place and particles of frequencies 
exceeding the mechanical frequency of the oscillating mirror are
created. The energy for this process is provided by the energy which
has to be given to the system from outside to maintain the motion of 
the mirror against the radiation reaction force \cite{Ford:1982,Barton:1993,
Maia Neto:1994,Mundarain:1998}.   
The more realistic case of a three-dimensional cavity is studied in 
\cite{Mundarain:1998,Crocce:2001, Dodonov:1995,Dodonov:2003,
Dodonov:2001,Crocce:2002,Dodonov:1998a,Dodonov:1998b}. 
Field quantization inside cavities with non-perfect boundary 
conditions has been investigated in, e.g., 
\cite{Schaller:2002a,Schaller:2002b} and corrections due to 
finite temperature effects are treated in 
\cite{Plunien:2000,Schuetzhold:2002,Jing:2000}. The question of how 
the quantum vacuum interacts with the (classical) dynamics of the
cavity has been addressed in 
\cite{Law:1994, Law:1995,Golestanian:1997,Cole:2001}.

In this work we present a formalism allowing for numerical investigation of the 
dynamical Casimir effect for scalar particles in a one-dimensional
cavity. (For related numerical work see also 
\cite{Antunes:2003,Li:2002,Fedotov:2005}.) 
We introduce a particular parametrization for the time evolution of
the field modes yielding a system of coupled first-order differential 
equations. The solutions to this system determine the number of
created particles and can be obtained by means of standard numerics. 
We employ the formalism to investigate the creation of real massless scalar 
particles in a resonantly as well as off-resonantly vibrating cavity 
and compare the numerical results with analytical predictions.
These results are complementary to the ones already presented in 
\cite{Ruser:2005}. 

With this formalism at hand the dynamical Casimir effect can be
investigated fully numerically making it possible to study a variety
of scenarios where no analytical results are known (large amplitude
oscillations, arbitrary wall motions etc.). Of special interest is 
of course the realistic case of the electromagnetic field in a 
three-dimensional cavity. Being easily extendable to arbitrary space 
dimensions the presented formalism can be used also in this case. In
particular it allows to calculate numerically the TE-mode contribution 
\cite{Crocce:2002} to the photon creation taking the influence of 
the intermode coupling fully into account \cite{Ruser:2005b}. 
Hence the formalism can be used to cross-check analytical results 
also in this realistic case which might be of importance for future 
experiments. Let us finally note that the tools (and the numerical 
formalism presented here) used to study the dynamical Casimir effect 
can also be employed to investigate graviton generation 
in braneworld cosmology \cite{Cartier:2005}.

\section{Hamiltonian and equations of motion}
We consider the Hamilton operator
\begin{equation}
\hat{H}(t)=\frac{1}{2}\sum_n\left[\hat{p}_n^2+\Omega_n^2(t)\hat{q}_n^2\right]
-\frac{1}{2}\sum_{nm}M_{nm}(t)\left[\hat{q}_n\hat{p}_m+
\hat{p}_m\hat{q}_n\right]
\label{Hamiltonian}
\end{equation}
with
\begin{equation}
\hat{p}_n=\dot{\hat{q}}_n+\sum_m\hat{q}_mM_{mn}(t)\;\;{\rm and}\;\;
M_{nm}(t)=\int_{I(t)}\,dx\,\dot{\phi}_n(t,x)\phi_m(t,x)
\end{equation}
describing the dynamics of a massless real scalar field $\Phi=\Phi(t,x)$ on a
time-dependent interval $I(t)=[0,l(t)]$ in terms of the canonical
operators $\hat{q}_n$ and $\hat{p}_n$ subject to the usual equal-time 
commutation relations. The corresponding canonical variables $q_n$ and
$p_n$ are introduced via the expansion
$\Phi(t,x)=\sum_n\,q_n(t)\phi_n(t,x)$ of the field and its momentum
$\Pi(t,x)=\sum_n\,p_n(t)\phi_n(t,x)$ in time-dependent (instantaneous)
eigenfunctions $\phi_n(t,x)$ satisfying the eigenvalue equation 
\begin{equation}
-\partial_x^2\phi_n(t,x)=\Omega_n^2(t)\phi_n(t,x)
\label{eigenvalue equation}
\end{equation}
on $I(t)$ with time-dependent eigenvalues $\Omega_n^2(t)$
\cite{Schuetzhold:1998}. The overdot denotes the derivative with 
respect to time $t$ and we are using units with $\hbar=c=1$. 
The time-dependent so-called coupling matrix
$M_{nm}$ arises due to the time-dependent boundary condition 
for the field at $x=l(t)$ enforcing the eigenfunctions of
$-\partial_x^2$ to be time-dependent. We take the boundary 
conditions at $x=0$ and $x=l(t)$ to be of the form
\begin{equation}
\left[a_1\Phi+a_2\partial_x\Phi\right]|_{x=0}=
\left[b_1\Phi+b_2\partial_x\Phi\right]|_{x=l(t)}=0\;\forall \;t
\label{boundary conditions}
\end{equation}
with constants $a_1,a_2,b_1$ and $b_2$ ensuring that the set of
eigenfunctions $\{\phi_n(t,x)\}$ is complete and orthonormal for all
times. 

Adopting the Heisenberg picture, the equations of motion for 
$\hat{q}_n(t)$ read \cite{Dodonov:1996,Schuetzhold:1998}
\begin{equation}
\ddot{\hat{q}}_n+\Omega_n^2\hat{q}_n+\sum_{m}
\left[M_{mn}-M_{nm}\right]\dot{\hat{q}}_m
+\sum_m\left[\dot{M}_{mn}-N_{nm}\right]\hat{q}_m=0          
\label{deq for q}
\end{equation}
where $N_{nm}=\sum_k M_{nk}M_{mk}$. The structure of the 
intermode coupling mediated by the coupling matrix $M_{nm}$ depends on the 
particular kind of boundary conditions which decide on the specific
form of the instantaneous eigenfunctions $\phi_n(t,x)$.
It is worth noting that the Hamiltonian (\ref{Hamiltonian}) does not 
correspond to the energy of the field because the coupling term does
not contribute to the total energy defined via the energy momentum
tensor \cite{Schuetzhold:1998}. From the Hamiltonian 
(\ref{Hamiltonian}) and the equations of motion (\ref{deq for q}) 
we identify two external time dependences in the equations 
which will lead to particle creation: (i) the time-dependent eigenfrequencies 
$\Omega_n(t)$ and (ii) the coupling matrix 
$M_{nm}(t)$, called the squeezing and acceleration effect, 
respectively.  

\section{Vacuum and particle definition}
Let us assume that the motion of the wall is switched on at $t=0$ with
$l(t)$ following a prescribed trajectory for a duration $t_1$, ceases 
afterwards and is at rest again. Before and after the motion 
the coupling matrix vanishes and the time evolution of the operator
$\hat{q}_n$ is determined by the equation of an harmonic oscillator
with constant frequency $\Omega_n^0 \equiv \Omega_n(t \le 0)$ and
$\Omega_n^1 \equiv \Omega_n(t \ge t_1)$, respectively
\footnote{Here the final position $l(t_1)=l_1$ is assumed to be 
arbitrary. In case of a vibrating cavity it is natural to consider 
times $t_1$ after which the dynamical wall has returned 
to its initial position.}.
The corresponding Hamilton operator describing the quantized field for 
$t\le 0$ and $t\ge t_1$ can then be diagonalized by introducing 
time-independent annihilation and creation operators 
$\{\hat{a}_n,\hat{a}_n^\dagger\}$, corresponding to the particle 
notion for $t\le 0$, and $\{\hat{A}_n,\hat{A}_n^\dagger\}$ 
associated with the particle notion
for $t\ge t_1$ via \footnote{We are assuming that $\Omega_n^0 \ne 0$ 
and $\Omega_n^1\neq 0$ for all $n$.}
\begin{eqnarray}
t\le0:\;\hat{q}_n(t)&=&
\frac{\hat{a}_n e^{-i\Omega_n^0\,t}}{\sqrt{2\Omega_n^0}}+
 {\rm h.c.},\;
\hat{p}_n(t)=i\sqrt{\frac{\Omega_n^0}{2}}
\hat{a}_n^\dagger\,e^{i\Omega_n^0t} +{\rm h.c.},
\label{reference q and p}\\
t\ge t_1:\;\hat{q}_n(t)&=&\frac{\hat{A}_n e^{-i\Omega_n^1\,(t-t_1)}}
{\sqrt{2\Omega_n^1}}+
 {\rm h.c.},\;
\hat{p}_n(t)=i\sqrt{\frac{\Omega_n^1}{2}}
\hat{A}_n^\dagger\,e^{i\Omega_n^1(t-t_1)} +{\rm h.c.}.
\label{final q and p}
\end{eqnarray}
The initial and final vacuum states $|0,t \le 0\rangle$ and $|0,t \ge t_1\rangle$,
respectively, are introduced as the ground states of the corresponding
diagonal Hamilton operators: 
\begin{equation}
\hat{H}=\sum_n\left\{
\begin{array}{l}
\Omega_n^0\,\left[\hat{a}_n^\dagger\hat{a}_n + 1/2\right]
\;\;\;{\rm with}\; \hat{a}_n\,|0,t \le 0\rangle=0
\;{\rm for} \; t \le 0\\
\\
\Omega_n^1\,\left[\hat{A}_n^\dagger\hat{A}_n + 1/2\right]
\;{\rm with} \;\hat{A}_n\,|0,t \ge t_1\rangle=0
\;{\rm for} \; t \ge t_1
\end{array} \right.
\label{reference hamiltonian}.
\end{equation} 
The set of initial-state operators $\{\hat{a}_n,\hat{a}_n^\dagger\}$
is related to the set of final-state operators 
$\{\hat{A}_n,\hat{A}_n^\dagger\}$ by a Bogoliubov transformation 
\begin{equation}
\hat{A}_n=\sum_m\left[{\cal A}_{mn}(t_1)\,\hat{a}_m+
{\cal B}_{mn}^*(t_1)\,\hat{a}_m^\dagger
\right]
\label{Bogoliubov transformation final particles}
\end{equation}
where ${\cal A}_{mn}(t_1)$ and ${\cal B}_{mn}(t_1)$ satisfy the relations
\begin{equation}
\sum_m\left[{\cal A}_{mn}{\cal A}^*_{mk}-{\cal B}^*_{mn}{\cal B}_{mk}
\right]=\delta_{nk}\;,\;
\sum_m\left[{\cal A}_{mn}{\cal B}^*_{mk}-{\cal B}^*_{mn}{\cal A}_{mk}
\right]=0.
\label{Bogoliubov relations}
\end{equation} 
For $t\ge t_1$ the particle number operator 
$\hat{N}_n=\hat{A}_n^\dagger\hat{A}_n$ 
defined with respect to the final vacuum state counts the number of 
physical particles. The number of particles created in a
mode $n$ during the motion of the wall is given as the expectation value of 
$\hat{N}_n$ with respect to the initial vacuum state $|0,t \le 0\rangle$:
\begin{equation}
N_n (t_1)= \langle0,t \le 0|\hat{N}_n|0,t \le 0\rangle =
\sum_m|{\cal B}_{mn}(t_1)|^2.
\label{phys particle number}
\end{equation}
Accordingly, total particle number $N(t_1)$ and energy $E(t_1)$ of
the motion induced radiation are given by
\begin{equation}
N(t_1)=\sum_n N_n(t_1)\;,\; E(t_1)=\sum_n \Omega_n^1 N_n(t_1).
\label{total particle number}
\end{equation}
Both quantities are in general ill defined and therefore require 
appropriate regularization. For a time dependence of the boundary 
$l(t)$ which is not sufficiently smooth, i.e. it exhibits discontinuities in 
its time-derivative appearing for instance when switching the motion 
on and off instantaneously, one may expect that part of the 
particle creation is due to this discontinuity in the velocity 
which may cause the excitation of modes of even arbitrary high 
frequencies. Hence a (large) contribution to the predicted 
particle creation may be spurious and in the case 
that arbitrary high frequency modes become excited  
the summations in (\ref{total particle number}) 
do not converge. This can be avoided most easily 
by introducing a frequency cut-off which effectively smoothes 
the dynamics $l(t)$. When calculating the quantities 
(\ref{total particle number}) numerically we will make use 
of such a frequency cut-off which is determined by the stability 
of the numerical results for single modes, i.e. stability of the 
expectation value (\ref{phys particle number}) with respect to
the cut-off. Note that an explicit frequency cut-off 
also accounts for imperfect (non-ideal) boundary conditions for 
high frequency modes \cite{Schuetzhold:1998}. 

\section{Time evolution}
During the motion of the boundary some or even infinitely many 
modes may be coupled. For $t\ge 0$ the operators 
$\hat{q}_n(t)$ and $\hat{p}_n(t)$ given by 
$\hat{U}^\dagger\hat{q}_n(0)\hat{U}$ 
and $\hat{U}^\dagger\hat{p}_n(0)\hat{U}$, respectively,
with $\hat{U}\equiv\hat{U}(t,0)={\cal T}\exp(-i\,\int_{0}^{t}\,dt'\hat{H}(t'))$
and ${\cal T}$ denoting the time-ordering operator, can be
expanded in initial state operators $\hat{a}_n$,
$\hat{a}^\dagger_n$ and complex functions $\epsilon_n^{(m)}(t)$:
\begin{eqnarray}
\hat{q}_n(t \ge 0)=\hat{U}^\dagger\hat{q}_n(0)\hat{U}=
\sum_m\frac{\hat{a}_m}{\sqrt{2\Omega_m^0}}
\;\epsilon_n^{(m)}(t) + {\rm h.c.}
\label{time evolution of q},\\
\hat{p}_n(t\ge0)=\hat{U}^\dagger\hat{p}_n(0)\hat{U}=
\sum_m\frac{\hat{a}_m}{\sqrt{2\Omega_m^0}}\left[
\dot{\epsilon}_n^{(m)}(t)+\sum_{k}M_{kn}(t)\epsilon_k^{(m)}(t)\right]+
{\rm h.c.}\quad
\label{time evolution of p}
\end{eqnarray}
By using the Heisenberg equation
$\dot{\hat{O}}(t)=i[\hat{H}(t),\hat{O}(t)]
+(\partial \hat{O}(t)/\partial t)_{\rm expl.}$ it is straightforward 
to show that the functions $\epsilon_n^{(m)}(t)$ satisfy the same 
differential equation (\ref{deq for q}) as $\hat{q}_n(t)$.
Notice that insertion of Eq.~(\ref{time evolution of q}) into the 
mode expansion for $\Phi$ leads to the decomposition of the field 
used in, e.g., \cite{Dodonov:1996,Ji:1997,Dodonov:1998}.   
Through the formal expansion (\ref{time evolution of q}) we have 
reduced the problem of finding the time evolution for the operator 
$\hat{q}_n(t)$ to the problem of solving the system of coupled 
second-order differential equations (\ref{deq for q}) for 
$\epsilon_n^{(m)}(t)$. Demanding that Eqs.~(\ref{time evolution of q}) 
and (\ref{time evolution of p}) have to match with the corresponding 
expressions (\ref{reference q and p}) at $t=0$ leads to the initial 
conditions
\begin{equation}
\epsilon_n^{(m)}(0)=\delta_{nm}\,,\quad 
\dot{\epsilon}_n^{(m)}(0)=-i\Omega_n^0\,\delta_{nm}-M_{mn}(0)
\label{inicond for epsilon}.
\end{equation}
Hence with $M_{mn}(0)$ vanishing only if $\dot{l}(0)=0$ the initial 
condition $\dot{\epsilon}_n^{(m)}(0)$ is not simply 
$-i\Omega_n^0\,\delta_{nm}$ when dealing with boundary 
motions $l(t)$ which have a discontinuity in the velocity at $t=0$.
Matching (\ref{final q and p}) with (\ref{time evolution of q}) and 
(\ref{time evolution of p}) at $t=t_1$ one finds 
\begin{equation}
{\cal A}_{mn}(t_1)=\frac{1}{2}\sqrt{\frac{\Omega_n^1}{\Omega_m^0}}
\,\left\{ \epsilon_n^{(m)}(t_1)+\frac{i}{\Omega_n^1}\left[
\dot{\epsilon}_n^{(m)}(t_1)+\sum_kM_{kn}(t_1)\epsilon_k^{(m)}(t_1)\right]
\right\}
\label{physical alpha}
\end{equation}
\begin{equation}
{\cal B}_{mn}(t_1)=\frac{1}{2}\sqrt{\frac{\Omega_n^1}{\Omega_m^0}}
\left\{\epsilon_n^{(m)}(t_1)-\frac{i}{\Omega_n^1}\left[
\dot{\epsilon}_n^{(m)}(t_1)+\sum_kM_{kn}(t_1)\epsilon_k^{(m)}(t_1)\right]
\right\}.
\label{physical beta}
\end{equation}
Starting with the initial vacuum $|0,t \le 0\rangle$ the Bogoliubov 
transformation (\ref{Bogoliubov transformation final particles}) has 
to become trivial for $t_1=0$, i.e. $\hat{A}_k=\hat{a}_k$, implying 
the vacuum initial conditions
\begin{equation}
{\cal A}_{mn}(0)=\delta_{mn}\quad {\rm and}\quad
{\cal B}_{mn}(0)=0 
\label{vac ini cond}
\end{equation}
which are consistent with the initial conditions 
(\ref{inicond for epsilon}). The emergence of $M_{mn}(0)$ in the
initial conditions (\ref{inicond for epsilon}) therefore 
guarantees to meet the vacuum initial conditions when the motion of 
the boundary starts instantaneously with a non-zero velocity. 

By introducing the auxiliary functions \footnote{A derivation can
be found in Appendix A.}
\begin{eqnarray}
\xi_n^{(m)}(t)&=&\epsilon_n^{(m)}(t)+\frac{i}{\Omega_n^0}\left[
\dot{\epsilon}_n^{(m)}(t)+\sum_k
M_{kn}(t)\,\epsilon_k^{(m)}(t)\right],
\label{def of small xi}\\
\eta_n^{(m)}(t)&=&\epsilon_n^{(m)}(t)-\frac{i}{\Omega_n^0}\left[
\dot{\epsilon}_n^{(m)}(t)+\sum_k M_{kn}(t)\,\epsilon_k^{(m)}(t)\right]
\label{def of small eta}
\end{eqnarray}
the expressions (\ref{physical alpha}) and (\ref{physical beta}) 
can be rewritten as
\begin{eqnarray}
{\cal A}_{mn}(t_1)&=&\frac{1}{2}\sqrt{\frac{\Omega_n^1}{\Omega_m^0}}
\left[\Delta_n^+(t_1)\,\xi_n^{(m)}(t_1)+
\Delta_n^-(t_1)\,\eta_n^{(m)}(t_1)
\right],
\label{relation between capital xi and eta and xi}\\
{\cal B}_{mn}(t_1)&=&\frac{1}{2}\sqrt{\frac{\Omega_n^1}{\Omega_m^0}}
\left[\Delta_n^-(t_1)\,\xi_n^{(m)}(t_1)+
\Delta_n^+(t_1)\,\eta_n^{(m)}(t_1)
\right]
\label{relation between capital eta and eta and xi}
\end{eqnarray} 
with 
\begin{equation}
\Delta_n^{\pm}(t)=\frac{1}{2}\left[1\pm\frac{\Omega_n^0}
{\Omega_n(t)}\right].
\label{big delta}
\end{equation}
The quantity $\Delta_n^{\pm}(t_1)$ is a measure for the 
deviation of the final state of the cavity, characterized by the 
cavity length $l(t_1)$, with respect to its initial size $l_0$. 
If at time $t_1$ the cavity size is equal
to the initial size $l_0$, for instance in the important case that 
$t_1$ is a multiple of the period of oscillations of the cavity, 
we have
${\cal B}_{mn}(t_1)=(1/2)\sqrt{\Omega_n^0/\Omega_m^0}\eta_n^{(m)}(t_1)$  
and therefore 
$N_n(t_1)=(1/4)\sum_m(\Omega_n^0/\Omega_m^0)|\eta_n^{(m)}(t_1)|^2$.

The advantage of introducing the functions $\xi_n^{(m)}$ and
$\eta_n^{(m)}$ is that they satisfy the following system of
first-order differential equations:
\begin{equation}
\dot{\xi}_n^{(m)}=-i\left[a^+_{nn}\xi_n^{(m)}-
a^-_{nn}\eta_n^{(m)}\right]
-\sum_k\left[c^-_{nk}\xi_k^{(m)}+c^+_{nk}\eta_k^{(m)}\right],
\label{deq for xi}
\end{equation}
\begin{equation}
\dot{\eta}_n^{(m)}=-i\left[a^-_{nn}\xi_n^{(m)}-
a^+_{nn}\eta_n^{(m)}\right]
-\sum_k\left[c^+_{nk}\xi_k^{(m)}+c^-_{nk}\eta_k^{(m)}\right]
\label{deq for eta}
\end{equation}
with
\begin{equation}
a_{nn}^\pm(t)=\frac{\Omega_n^0}{2}
\left\{1 \pm \left[\frac{\Omega_n(t)}{\Omega_n^0}\right]^2\right\}
\;{\rm and}\;c_{nk}^\pm(t)=\frac{1}{2}\left[M_{kn}(t) \pm 
\frac{\Omega_k^0}{\Omega_n^0}M_{nk}(t)\right].
\label{def a matrix}
\end{equation}
Besides the time-dependent frequency $\Omega_n(t)$ only the 
coupling matrix $M_{kn}(t)$ enters into this system of coupled 
differential equations but neither $N_{nk}(t)$ nor its time derivative 
$\dot{M}_{kn}(t)$. The vacuum initial
conditions (\ref{vac ini cond}) entail the initial conditions
for the functions $\xi_n^{(m)}$ and $\eta_n^{(m)}$ to be
\begin{equation}
\xi_n^{(m)}(0)=2\delta_{mn}\;,\;\;\;\eta_n^{(m)}(0)=0.
\label{initial conditions for xi and eta}
\end{equation}
Let us stress that all derivations and equations shown so far, 
do not rely on particular symmetry properties of the coupling matrix.   

By means of Eq.~(\ref{relation between capital eta and eta and xi}) 
the number of particles created from vacuum during the dynamics of the
cavity as well as the associated energy may now be calculated 
from the solutions $\xi_n^{(m)}$ and 
$\eta_n^{(m)}$ of the system of coupled first-order differential equations
formed by Eqs.~(\ref{deq for xi}) and (\ref{deq for eta}). 

In order to obtain the numerical results presented in the next section 
we proceed in the following way: A cut-off parameter 
$k_{\rm max}$ is introduced to make the system of differential 
equations finite and suitable for a numerical treatment. 
The system of coupled differential equations is then 
evolved numerically from $t=0$ up to a final time $t_{\rm max}$ 
and the expectation value (\ref{phys particle number}) 
is calculated for several times  in between. 
By doing so we interpret $t_1$ as a continuous variable 
such that Eq.~(\ref{phys particle number}) becomes a continuous 
function of time\footnote{Interpreting $t_1$ as a continuous function
of time one can of course derive a corresponding system of 
coupled differential equations for ${\cal A}_{mn}$ and 
${\cal B}_{mn}$ (see Appendix B).}. 
Consequently, the stability of the numerical solutions with respect to the cut-off  
has to be ensured. In particular $k_{\rm max}$ will be chosen 
such that the numerical results for the number of particles created in 
single modes (\ref{phys particle number}) are stable. 
Furthermore, the quality of the numerical results is assessed by
testing the Bogoliubov relations (\ref{Bogoliubov relations}).     
  
This procedure is of course not without problems when the 
expectation values are evaluated also for times $t_1$ at which 
$\dot{l}(t_1)\neq 0$. The used particle definition requires 
then a matching of the solutions to expressions corresponding 
to the static configuration with $\dot{l}(t_1)=0$, 
hence a discontinuity in the velocity appears which may cause 
spurious effects. However the cut-off automatically ensures that possible
spurious effects do not yield a 
divergent total particle number (see also section 3).
Indeed we will see that in the particular scenario of interest 
- vibrating cavity - the influence of this matching problem
is tiny and the numerical results agree perfectly with analytical
predictions.  

\section{Numerical results and discussion}
In this section we consider a massless real scalar field 
subject to Dirichlet boundary conditions at $x=0,l(t)$ 
and the much studied sinusoidal cavity motion 
\begin{equation}
l(t)=l_0\left[1+\epsilon\sin(\omega t)\right]\;,\;\;\epsilon \ll 1.
\label{sine motion}
\end{equation}
The time-dependent frequency and coupling matrix are given 
by \cite{Schuetzhold:1998}
\begin{equation}
\Omega_n(t)=\frac{n\pi}{l(t)}\;,\;\;M_{nm}=\frac{\dot{l}(t)}{l(t)}
(-1)^{n+m}\frac{2nm}{m^2-n^2}
\end{equation}
for $n\neq m$ and $M_{nn}(t)=0$ with $n,m=1,2,3,...$ . 
The motion (\ref{sine motion}) whose absolute value of the 
velocity is maximal at the beginning of the motion as well as 
for times at which the wall returns to its initial position 
features the above described matching problem.
In \cite{Ruser:2005} we have already studied particle creation caused
by this motion for the main resonance case $\omega=2\Omega_1^0$ 
with the same formalism. We have found that for sufficiently 
small $\epsilon$ and appropriate $k_{\rm max}$ 
the numerical results are in excellent agreement with analytical 
predictions of \cite{Dodonov:1996} and \cite{Dodonov:1993}. 
Furthermore, the influence of the initial discontinuity in the
velocity of the motion (\ref{sine motion}) has been investigated 
showing that it is negligible for $\epsilon \ll 1$.  
 
Here we want to concentrate on higher resonances $\omega=2\Omega_n^0$ 
with $n>1$ and off-resonant frequencies (detuning). 
In the simulations we set $l_0=1$ and $\epsilon=0.001$. 
For these parameters it is shown in \cite{Ruser:2005} that the numerical 
results agree very well with analytical predictions derived under the assumption 
$\epsilon \ll 1$. The numerical results are compared with 
analytical expressions obtained in \cite{Ji:1997,Dodonov:1998}.
Remarks about the numerics can be found in Appendix C.

In Fig.~\ref{figure 1} (a) we show the numerical results for the total
particle number in the time range $[0,250]$ for resonant cavity 
frequencies $\omega=2\Omega_n^0=2n\pi$ for $n=1.5,2,2.5$ and 
$3$ and the associated energy of the created quantum radiation 
is depicted in Fig.\ref{figure 1} (b). The corresponding particle
spectra are shown in Fig.~\ref{figure 2} for different cut-off
parameters $k_{\rm max}$ to demonstrate numerical stability of the
results. Here stability of the numerical results means that
for the lowest modes $k$ the value $N_k(t)$ remains unchanged 
(within numerical precision) under variation of $k_{\rm max}$. 
The spectra confirm that no modes $k=2np$ 
with $p=1,2,3,...$ are coupled (and therefore excited) as predicted 
by the coupling condition $\omega=|\Omega_k^0 \pm \Omega_l^0|$ 
derived and discussed in \cite{Crocce:2001}.  
\begin{figure}
\begin{center}
\begin{tabular}{cc}
\includegraphics[height=5cm]{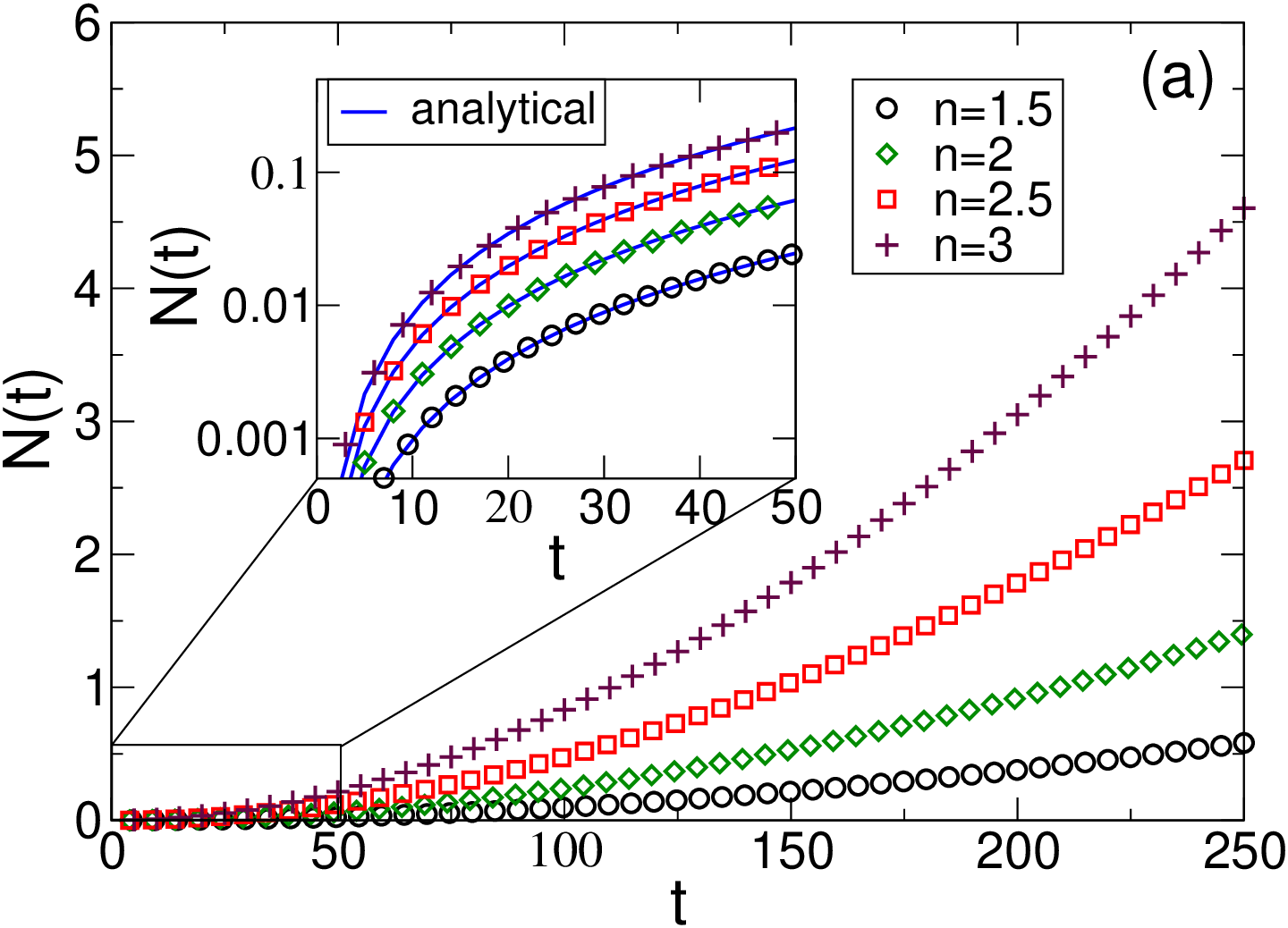}
&
\includegraphics[height=5cm]{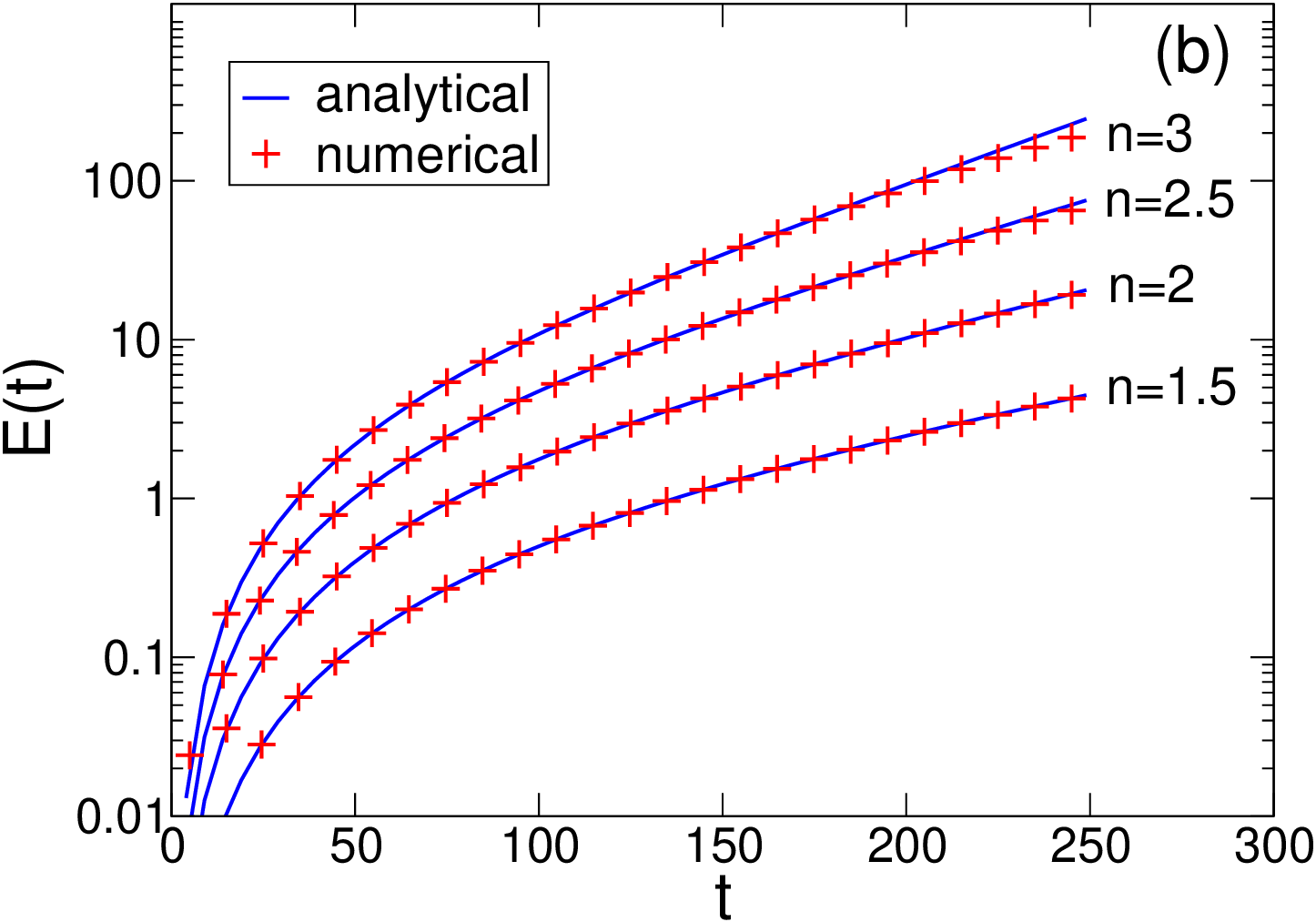}
\end{tabular}
\caption{(a) Total number of particles produced in a cavity vibrating with 
(\ref{sine motion}) and $\omega=2n\pi$ with $n=1.5,2,2.5$ and $3$.
The small plot shows the results in the time range $[0,50]$ 
together with the analytical prediction (solid line)
$N(t)=n(4n^2-1)(10^{-3}\pi t)^2/12$ of
\cite{Ji:1997,Dodonov:1998} valid for short times $(10^{-3}\pi t)\ll 1$.
(b) Numerical results for the created energy $E(t)$ corresponding to
(a) together with the analytical prediction
$E(t)=(4n^2-1)\pi\sinh^2(n10^{-3}\pi t)/12$ of \cite{Dodonov:1998}
(solid line). The results correspond to the largest cut-off parameters 
as given in Fig. \ref{figure 2}.
\label{figure 1}}
\end{center}
\end{figure}

\begin{figure}
\begin{center}
\begin{tabular}{cc}
\includegraphics[height=5cm]{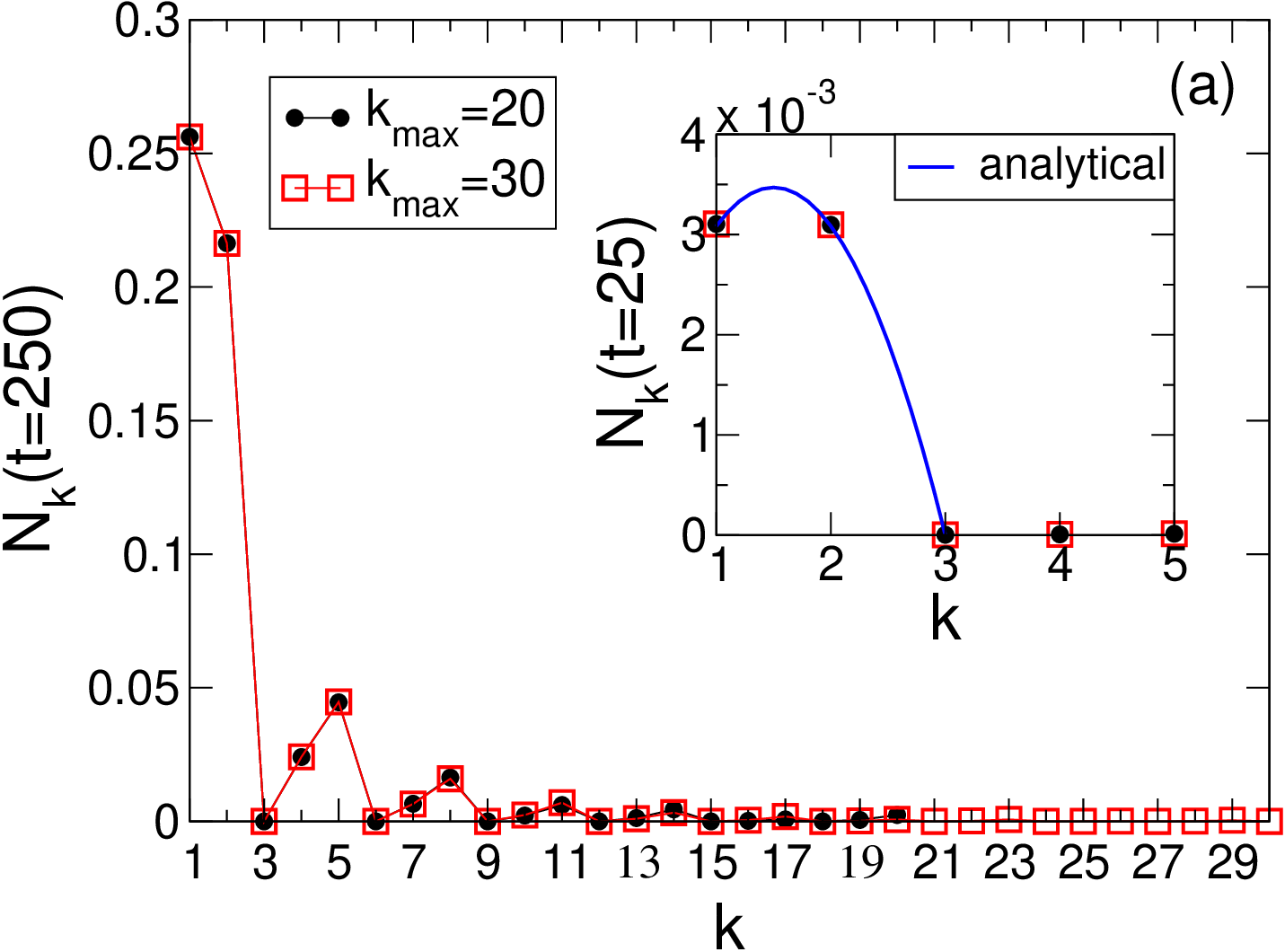}
&
\includegraphics[height=5cm]{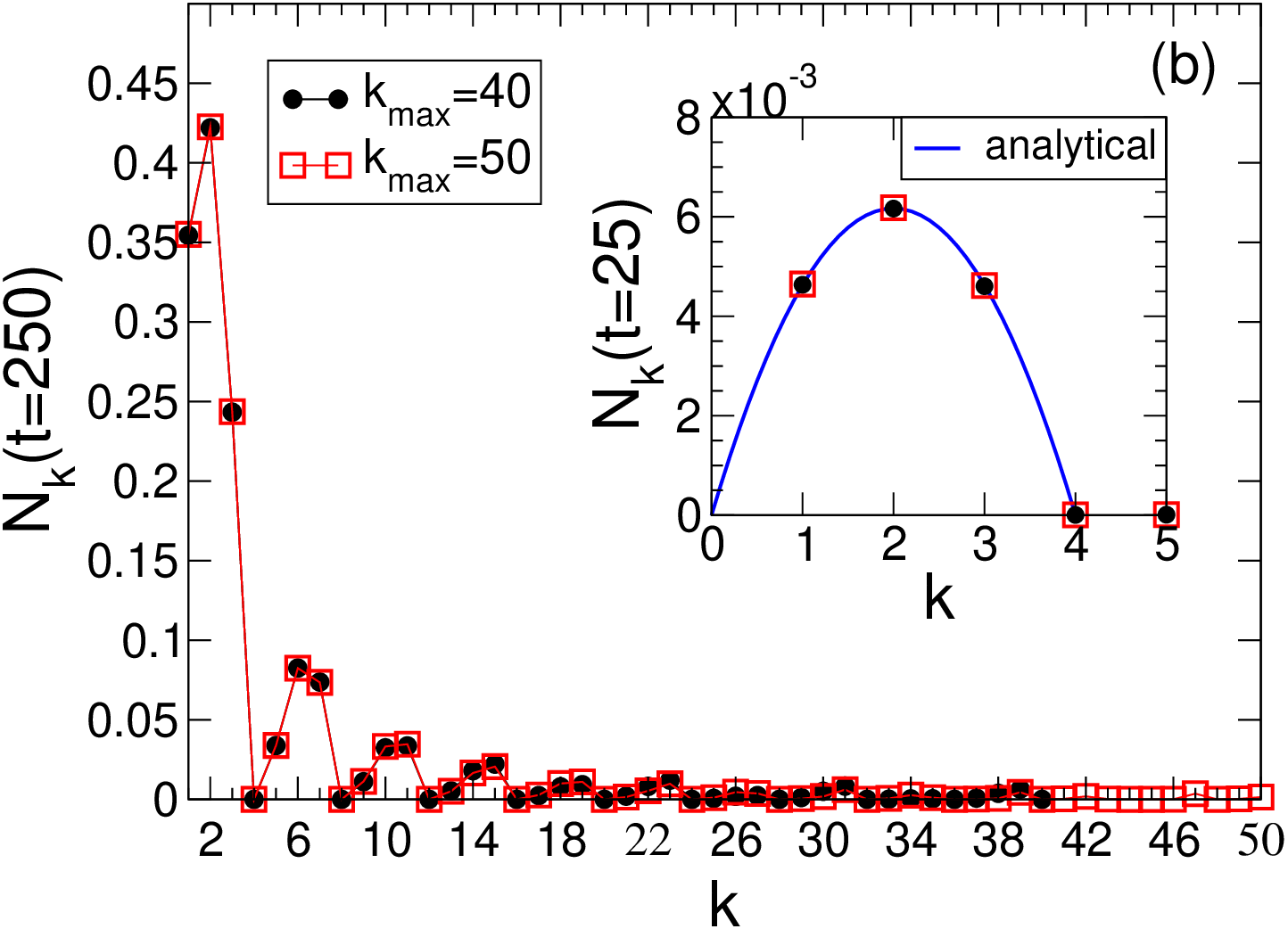}
\\
\includegraphics[height=5cm]{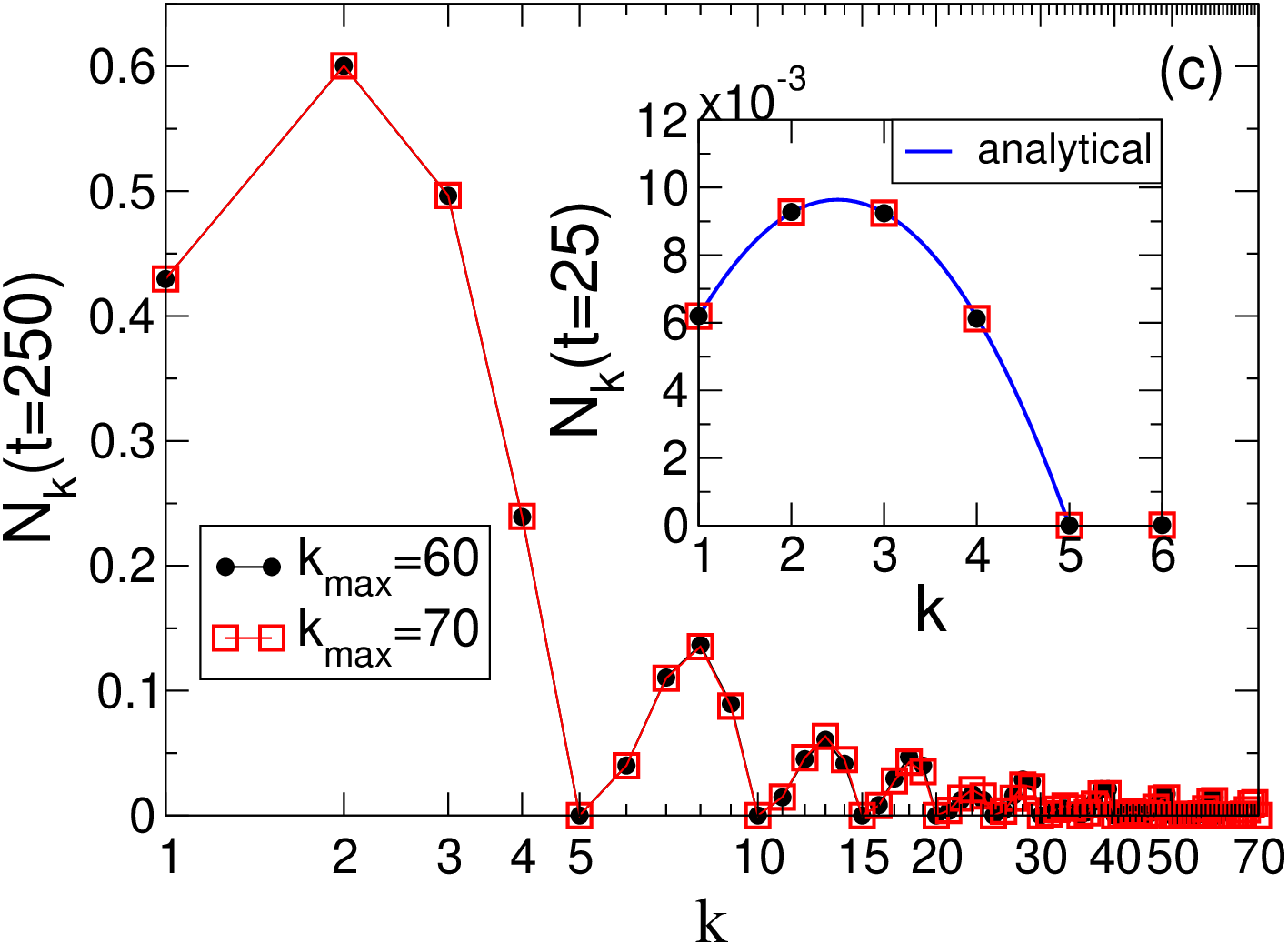}
&
\includegraphics[height=5cm]{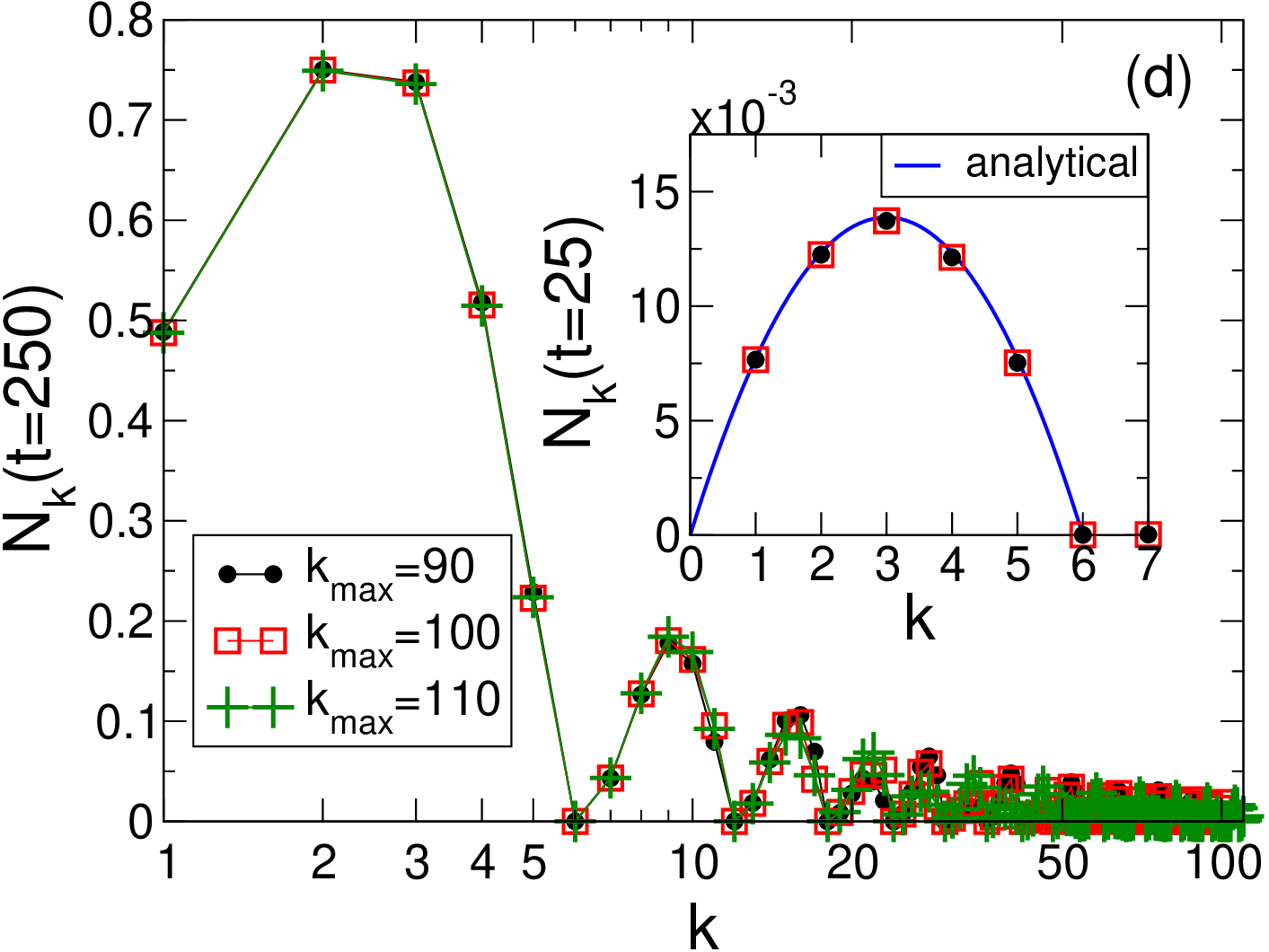}
\end{tabular}
\caption{Particle spectra for (a) $\omega=3\pi$, (b) $\omega=4\pi$, 
(c) $\omega=5\pi$ and (d) $\omega=6\pi$ corresponding to the results 
shown in Fig. \ref{figure 1}. The small plots compare the numerical
results for $N_k(t=25)$ with the analytical prediction 
$N_k(t)=(2n-k)k(10^{-3}\pi t)^2/4$ of \cite{Ji:1997} plotted for
continuous values of $k$ (solid line).
\label{figure 2}}
\end{center}
\end{figure}

For short times $\epsilon\,\pi\,t = 10^{-3}\,\pi\,t \ll 1$ the numerical results 
are well described by the analytical predictions of 
\cite{Ji:1997,Dodonov:1998}. The numerically calculated
spectra for times $t=25$ shown in Fig.~\ref{figure 2} are well 
fitted by the analytical expression 
$N_k(t)=(2n-k)k(10^{-3}\pi t)^2/4$ for $k < 2n$ and
$N_k(t)=0$ otherwise \cite{Ji:1997}, predicting a parabolic shape of the 
particle spectrum. More quantitatively, for $n=2$, for instance, the 
predicted values $N_1(t=25)=N_3(t=25)\sim 4.63\times 10^{-3}$,
$N_2(t=25)=6.17\times 10^{-3}$ agree well with the 
values $N_1(t=25)=4.62 \times 10^{-3}$, $N_2(t=25)=6.14 \times 10^{-3}$
and $N_3(t=25)=4.59 \times 10^{-3}$ obtained from the numerical 
simulations with $k_{\rm max}=50$. Consequently, the total number 
of created particles is perfectly described by the expression 
$N(t)=n(4n^2-1)(10^{-3}\pi t)^2/12$  \cite{Ji:1997,Dodonov:1998} as it is 
demonstrated in the small plot in Fig. \ref{figure 1} (a).

For the entire integration range $[0,250]$ we compare the 
numerical results for the total energy associated with the created
quantum radiation with the  analytical expression 
$E(t)=(4n^2-1)\pi\sinh^2(n\,10^{-3}\pi t)/12$  
\cite{Dodonov:1998} predicting that the energy increases exponentially
with time [Fig. \ref{figure 1} (b)]. 
The numerical values and the analytical prediction agree very well 
for $n=1.5$ and $2$. In the case of $n=2.5$ and $3$ we observe 
slight deviations towards the end of the integration range. 
This is due to the numerical instabilities in the corresponding
particle spectra [cf Figs. \ref{figure 2} (c) and (d)]. 
The numerical values for $N_k$ with $k$ larger than some value
($k > 10$ for $n=3$, for instance) do not remain unchanged when 
varying $k_{\rm max}$. Even $N_k$ is small for the higher
frequencies compared to the values of $N_k$ for the excited lowest 
modes their contribution to the total energy is significant 
because of their high frequency. Hence relatively small instabilities 
in $N_k$ for larger $k$ give rise to a non-stable 
(with respect to $k_{\rm max})$ result for the energy. 
In order to gain better agreement of the numerical results 
for the energy for $n=2.5$ and $3$ with the analytical prediction
a further increase of $k_{\rm max}$ is necessary.  

\begin{figure}
\begin{center}
\begin{tabular}{cc}
\includegraphics[height=5cm]{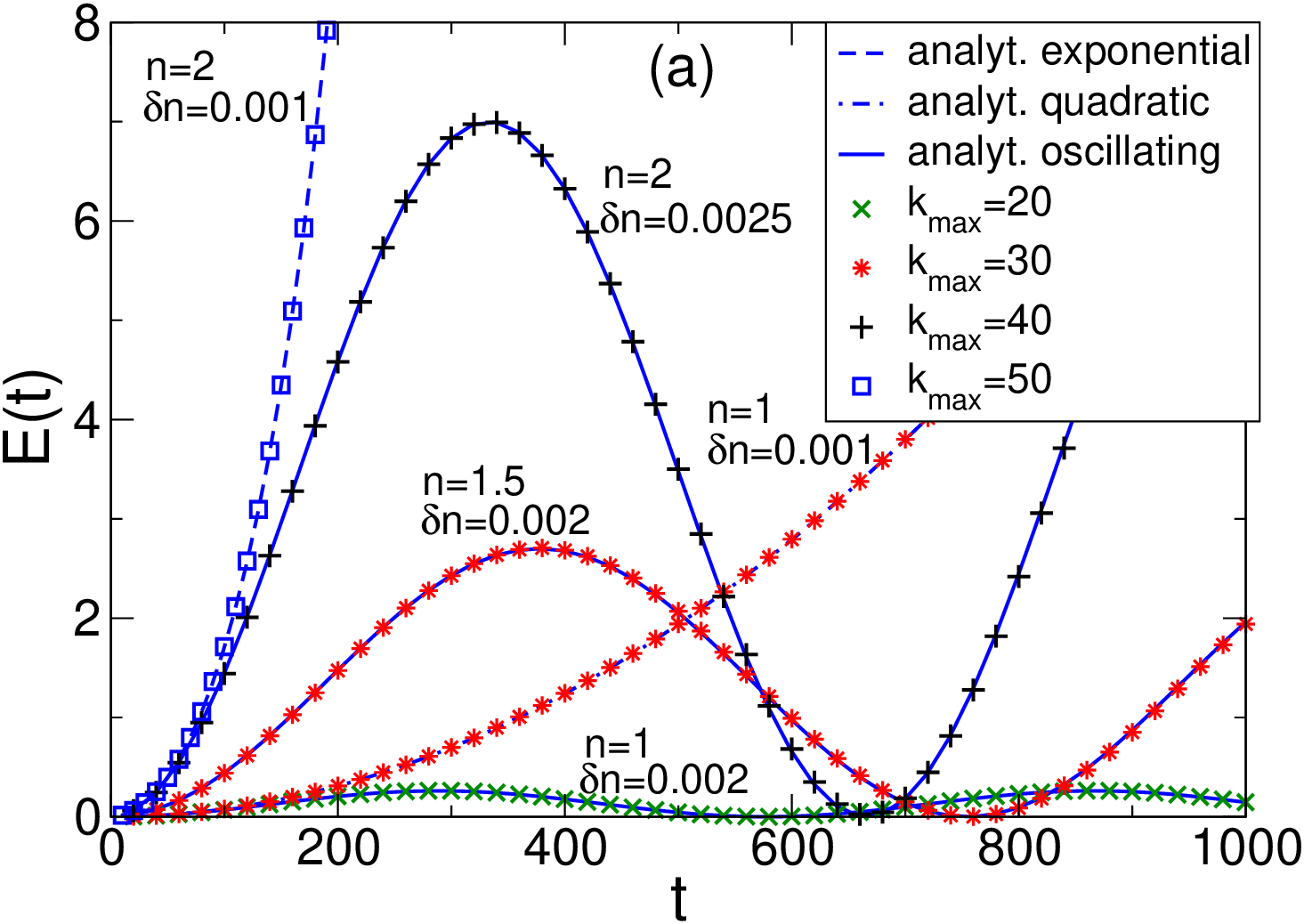}
&
\includegraphics[height=5cm]{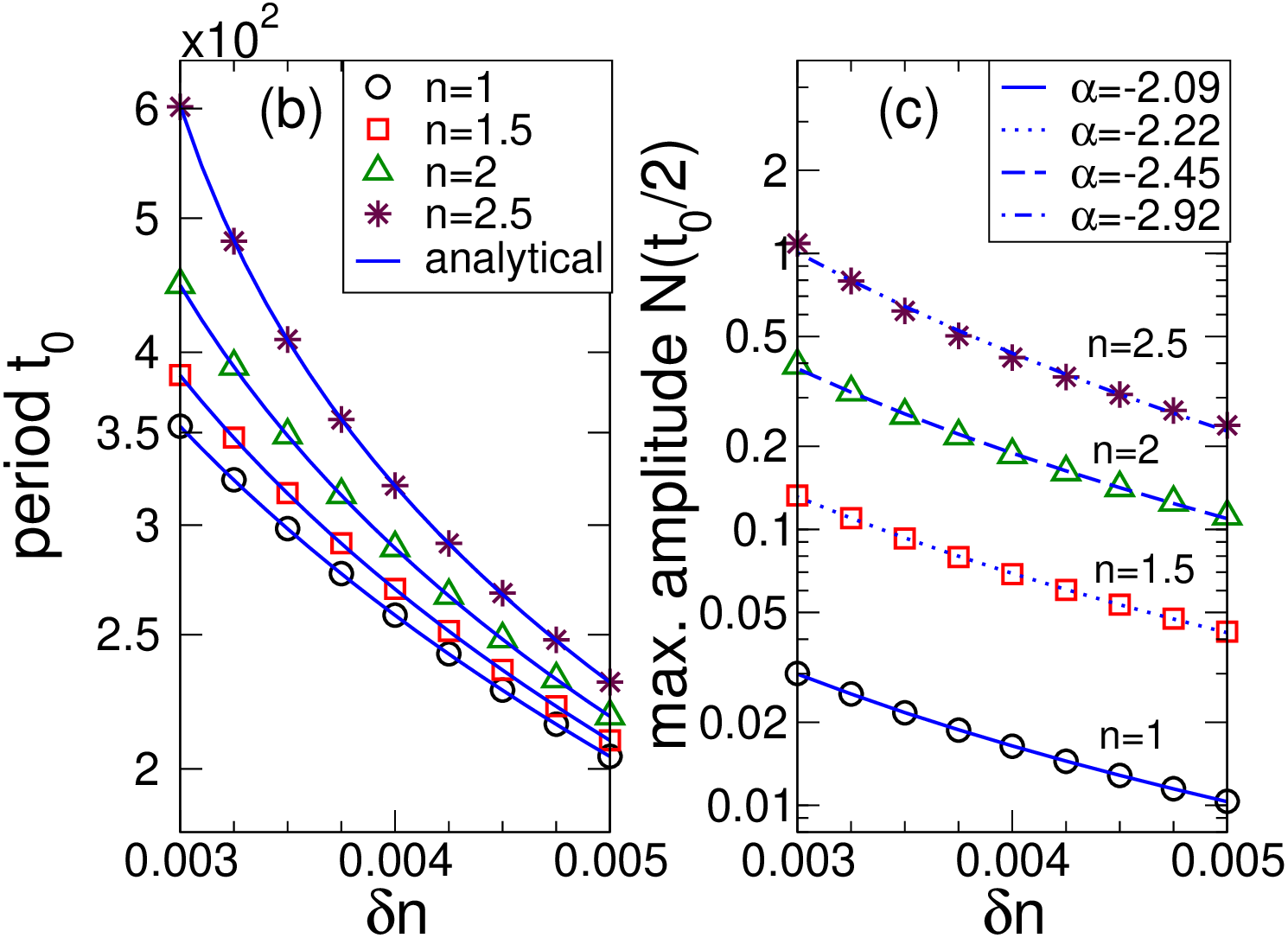}
\end{tabular}
\caption{(a) Energy associated with the particles created in an 
  off-resonantly vibrating one-dimensional
  cavity. Numerical results are compared to the analytical predictions
  of \cite{Dodonov:1998}. The numerical results are always shown for
  cut-off parameters $k_{\rm max}$ which ensure numerical stability.
  (b) Period of particle number oscillations caused by detuning. 
  The numerically obtained period is compared with the analytical prediction
  $t_0=10^{3}/(n\sqrt{\gamma^2-1})$ of \cite{Dodonov:1998}.
  (c) Maximal amplitude $N(t_0/2)$ of the corresponding particle
   number oscillations fitted to the power law 
   $N(t_0/2)\propto (\delta n)^\alpha$.
\label{figure 3}}
\end{center}
\end{figure}

We now consider the case of detuning $\omega=2\pi(n+\delta n)$. 
In an off-resonant vibrating one-dimensional cavity the total 
energy associated with the created particles may increase
exponentially 
$E(t)=(\pi/12)(4n^2-1)\sinh^2(n\sqrt{1-\gamma^2}\pi10^{-3}\,t)/(1-\gamma^2)$ if 
$\gamma<1$, quadratically $E(t)=(\pi/3)(4n^2-1)(n\pi10^{-3}\,t/2)^2$
if $\gamma=1$ or oscillate 
$E(t)=(\pi/12)(4n^2-1)\sin^2(n\sqrt{\gamma^2-1}\pi10^{-3}\,t)/(\gamma^2-1)$
if $\gamma>1$, depending on the strength of detuning 
$\delta n$ parametrized by $\gamma=\delta n \times 10^{3}/n$ 
\cite{Dodonov:1998}. In Fig.~\ref{figure 3} (a) results for the total
energy obtained in simulations with different off-resonant frequencies 
are shown covering all three different possibilities for $\gamma$ and compared 
to the analytical predictions. In all cases the numerical
results are very well described by the analytical expressions. 
Figure \ref{figure 3} (b) depicts the periods of the energy 
(and particle number) oscillations as obtained from the simulations and
compares them with the analytical prediction
$t_0=10^3/(n\sqrt{\gamma^2-1})$ showing that both are in good agreement. 
The numerical values for the maximal amplitudes $N(t_0/2)$ of the 
corresponding particle number oscillations are shown in Fig. 
\ref{figure 3} (c) and fitted to the power law $N(t_0/2) \propto
(\delta n)^\alpha$ with values of $\alpha$ as indicated in the figure.

The numerical results presented in this section are entirely in very 
good agreement with the corresponding analytical predictions derived
for small amplitude oscillations $\epsilon \ll 1$ which 
demonstrates the reliability of the numerical simulations. 
However, a few critical comments are in order. In our considerations
above the analytical expressions have been treated as continuous
functions of time. But strictly speaking, they are valid only for times
at which the moving wall has returned to its initial position.   
Moreover, in the numerical simulations the expectation values 
have been calculated also for times at which the velocity of the moving wall 
is non-zero (matching problem). Consequently one may
expect that part of the particle production is spurious and in
particular if modes of arbitrarily high frequency are excited, 
for instance due to the initial discontinuity
in the velocity of the motion (\ref{sine motion}), and 
$k_{\rm max} \rightarrow \infty$ the particle number
diverges and the numerical results do not
agree with the analytical predictions. In the numerical simulations
this is automatically avoided due to the cut-off $k_{\rm max}$.  
Nevertheless, working with a finite cut-off is well motivated because
it simulates imperfect boundary conditions for high frequency
modes and, as a matter of course, is a necessity for 
a numerical treatment. For the numerical results presented 
above these spurious effects are negligibly small and therefore the
numerical results agree very well with the analytical predictions. 
This is due to the fact that we restrict ourselves to small 
amplitudes $\epsilon \ll 1$ and hence to small velocities. 
Therefore the effect of the discontinuity in the velocity of the
boundary motion on the particle creation is expected to be small. 
This has been studied for the initial
discontinuity in the velocity in \cite{Ruser:2005} in detail 
demonstrating that the matching problem becomes only important 
for larger $\epsilon$. In Appendix C we discuss the
convergence of the numerical results in comprehension with 
the numerical accuracy of the simulations for the 
case $\omega=3\pi$ more detailed. 
 
\section{Conclusion}
A formalism allowing for numerical investigation of 
particle creation from vacuum in dynamical cavities, i.e. the
dynamical Casimir effect, has been presented. 
By introducing a particular parametrization 
for the time-evolution of the field modes inside the dynamical cavity, 
a system of coupled first-order linear differential
equations has been derived. Physical quantities like the number of 
particles created during the dynamics of the cavity and the associated 
energy are determined by the solutions 
to this system which can be found by applying standard numerics.  

In continuation of the work \cite{Ruser:2005} we have studied the creation of 
massless scalar particles due to resonant as well as off-resonant 
sinusoidal oscillations of one of the cavity walls. 
The numerical results are entirely in agreement with the 
analytical predictions derived in \cite{Ji:1997,Dodonov:1998}
demonstrating that the numerical simulations are reliable and 
the method introduced is appropriate to study the dynamical Casimir 
effect fully numerically.
 
Potential problems inherent in the method, in particular
the matching problem due to discontinuities in the velocity of the
boundary motion yielding spurious contributions to the 
total particle number, have been discussed. It has been shown that 
this effect is negligibly small for cavity vibrations with a sufficiently 
small amplitude.  

Being derived very generally, the method is applicable for different 
kinds of boundary conditions of the form (\ref{boundary conditions}) 
provided that the spectrum $\{\Omega_n\}$ contains no zero mode, 
i.e. $\Omega_n > 0\;\forall n$ and can easily be extended to massive
scalar fields by substituting for the frequency $\Omega_n$ the 
corresponding expression for a massive scalar field, 
i.e. $\Omega_n=\sqrt{(n\pi)^2/l^2+m^2}$ where $m$ is the mass. 
Furthermore, the generalization to higher
dimensional cavities is straightforward. This makes it possible 
to study the dynamical Casimir effect for a variety of possible 
interesting scenarios where less or even nothing is known analytically.
As already mentioned in the introduction, TE-mode photon creation 
in a three-dimensional rectangular cavity \cite{Crocce:2001} 
can be studied with the same method as well because it can be 
related to the production of massive scalar particles in 
a one-dimensional cavity \cite{Ruser:2005b}. However, more complicated 
boundary conditions than (\ref{boundary conditions}) 
appearing for example when studying TM-mode photons \cite{Crocce:2002}
cannot be treated within this approach.

\ack
The author would like to thank Ruth Durrer for comments on the
manuscript as well as Diego Dalvit and Francisco Mazzitelli for 
discussions during the QFEXT'05 workshop and Emilio Elizalde 
for organizing it. Financial support from the Swiss National 
Science Foundation is gratefully acknowledged.
\begin{appendix}

\section{Derivation of $\xi_n^{(m)}$ and $\eta_n^{(m)}$}
The auxiliary functions $\xi_n^{(m)}(t)$ and $\eta_n^{(m)}(t)$ 
[Eqs.~(\ref{def of small xi}), (\ref{def of small eta})]
can be introduced in the following way: Define an operator   
$\hat{b}_n(t)$ via $\hat{b}_n(t):=\hat{U}^\dagger(t,0)\,
\hat{a}_n\,\hat{U}(t,0)$ with $\hat{a}_n$ being the annihilation
operator corresponding to the initial state 
[Eq.~(\ref{reference q and p})], i.e. 
$\hat{a}_n=(1/\sqrt{2})[\sqrt{\Omega_n^0} \hat{q}_n(0)+
(i/\sqrt{\Omega_n^0})\hat{p}_n(0)]$. Hence 
$\hat{b}_n(t)=(1/\sqrt{2})[\sqrt{\Omega_n^0} \hat{U}^\dagger(t,0)
\hat{q}_n(0)\hat{U}(t,0)+(i/\sqrt{\Omega_n^0})
\hat{U}^\dagger(t,0)\hat{p}_n(0)\hat{U}(t,0)]$. 
By using Eqs. (\ref{time evolution of q}) and 
(\ref{time evolution of p}) one derives 
\begin{equation}
\hat{b}_n(t \ge 0)=\sum_m\frac{1}{2}\sqrt{\frac{\Omega_n^0}{\Omega_m^0}}
\left[\xi_n^{(m)}(t)\;\hat{a}_m+\eta_n^{(m)^*}(t)\;\hat{a}_m^\dagger\right]
\label{time evolution of initial op}
\end{equation}
with $\xi_n^{(m)}(t)$ and $\eta_n^{(m)}(t)$ defined in Eqs. 
(\ref{def of small xi}) and (\ref{def of small eta}).
Note that this definition of the time evolution for $\hat{b}_n(t)$ 
does not account for an explicit time-dependence of 
$\hat{b}_n(t)$. Therefore, in general, no meaningful notion of 
particles may be assigned to the operator $\hat{b}_n(t)$. 
This manifests itself in the relation between the operator 
$\hat{A}_n$ corresponding to the 
particle notion for $t\ge t_1$ [Eq. (\ref{final q and p})] 
and the operators $\hat{b}_n(t_1)$, $\hat{b}^\dagger_n(t_1)$ given by
\begin{equation}
\hat{A}_n=\sqrt{\frac{\Omega_n^1}{\Omega_n^0}}
\left[\Delta_n^+(t_1)\,\hat{b}_n(t_1)+\Delta_n^-(t_1)\,\hat{b}_n^\dagger(t_1)
\right]
\label{relation between b and big a}
\end{equation}
with $\Delta_n^{\pm}(t)$ defined in (\ref{big delta}). Equation 
(\ref{relation between b and big a}) follows directly
from Eq.(\ref{Bogoliubov transformation final particles}) 
with (\ref{relation between capital xi and eta and xi}), 
(\ref{relation between capital eta and eta and xi}) 
and (\ref{time evolution of initial op}). For motions ending 
at $t=t_1$ with $l(t_1)\neq l_0 $ the operator $\hat{b}_n(t)$ 
has not evolved into the operator $\hat{A}_n$ associated with the 
correct particle notion after the dynamics. However, if $l(t_1)=l_0$,
for example when $t_1$ is a multiple of the period of boundary
vibrations, $\hat{A}_n=\hat{b}_n(t_1)$.     

\section{The system for ${\cal A}_n^{(m)}$ and ${\cal B}_n^{(m)}$}
Taking the stopping time $t_1$ in (\ref{physical alpha}) and 
(\ref{physical beta}) to be a continuous variable one derives 
the following system of coupled differential equations for 
${\cal A}_{mn}$ and ${\cal B}_{mn}$:
\begin{eqnarray}
\dot{\cal A}_{mn}=-i\Omega_n{\cal A}_{mn}+\Gamma_n{\cal B}_{mn}
+\sum_k\left[K_{nk}^-{\cal A}_{mk} - K_{nk}^+{\cal B}_{mk}\right]\\
\dot{\cal B}_{mn}=-i\Omega_n{\cal B}_{mn}+\Gamma_n{\cal A}_{mn}
+\sum_k\left[K_{nk}^-{\cal B}_{mk} - K_{nk}^+{\cal A}_{mk}\right]
\end{eqnarray}
with 
\begin{equation}
\Gamma_n(t)=\frac{1}{2}\frac{\dot{\Omega}_n(t)}{\Omega_n(t)}\;,\;\;
K_{nk}^\pm(t)=\frac{1}{2}\left[\sqrt{\frac{\Omega_k(t)}{\Omega_n(t)}}M_{nk}(t)\pm
\sqrt{\frac{\Omega_n(t)}{\Omega_k(t)}}M_{kn}(t)\right].
\end{equation}

\section{Numerics}
\begin{center}
\begin{figure}[h!]
\begin{minipage}[c]{3.1in}
\includegraphics[height=5cm]{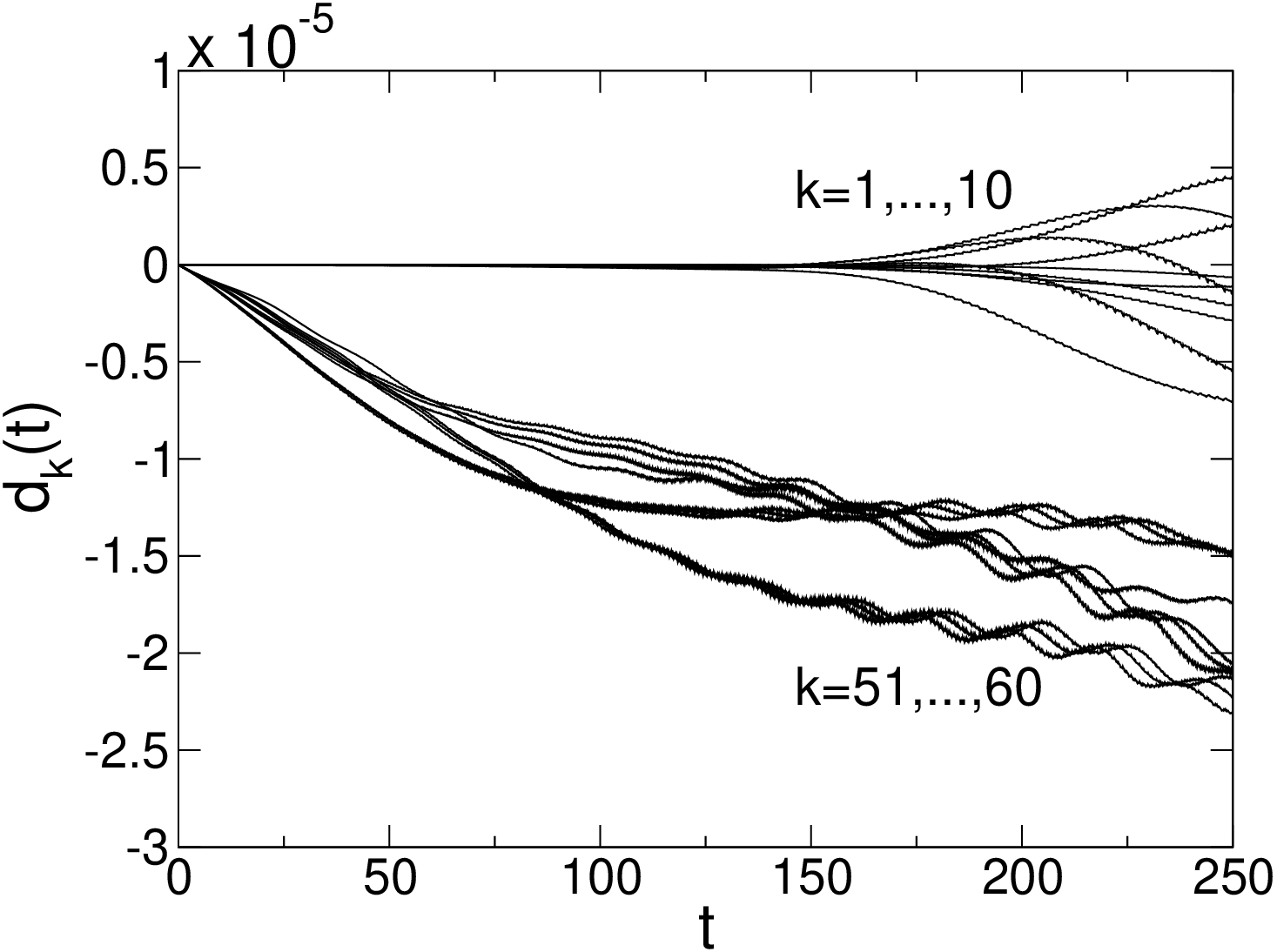}
\end{minipage}
\begin{minipage}[c]{3.2in}
\begin{tabular}{|c|c|c|}
\hline
$k_{\rm max}$ & $N(t=249.5)$ & $N(t=250.0)$\\\hline
20 & 0.5799007 & 0.5823052 \\ 
30 & 0.5798943 & 0.5822980 \\
40 & 0.5798951 & 0.5822983 \\
50 & 0.5798956 & 0.5822984 \\
60 & 0.5798959 & 0.5822984 \\\hline
\end{tabular}
\end{minipage}
\caption{Left: Plot showing the numerically evaluated diagonal part of the
  first of the Bogoliubov relations (\ref{Bogoliubov relations}) 
  $d_k=1-\sum_m(|{\cal A}_{mk}|^2-|{\cal B}_{mk}|^2)$ for 
  the lowest frequencies $k=1,...,10$ as well as $k=51,...,60$
  corresponding to the simulation with cut-off $k_{\rm max}=60$.
  Right: Table showing the numerical values of the total
  particle number at times $t=249.5$ and $t=250$ obtained 
  for $\omega=3\pi$ and $k_{\rm max}=20,30,40,50$ and $60$. 
\label{figure 4}}
\end{figure}
\end{center}

The numerical simulations have been performed by using 
a Runge-Kutta Prince-Dormand method ({\tt rk8pd}) based on 
source code provided by the GNU Scientific Library (GSL) \cite{gsl}. 
In the table in Fig. \ref{figure 4} we show the numerical values for the total
number of particles $N(t)$ created in a cavity subject to sinusoidal  
oscillations of the form (\ref{sine motion}) with frequency 
$\omega=3\pi$ [cf. Figs. \ref{figure 1} and \ref{figure 2} (a)] 
for two times $t=249.5$ and $t=250.0$ and cut-off parameters 
$k_{\rm max}=20,30,40,50$ and $60$. The plot in Fig. \ref{figure 4} 
shows the diagonal part of the first of the Bogoliubov relations 
(\ref{Bogoliubov relations}) 
$d_k=1-\sum_m(|{\cal A}_{mk}|^2-|{\cal B}_{mk}|^2)=0$ 
for $k=1,...,10$ and $51,...,60$ computed from the solutions 
of the simulation with $k_{\rm max}=60$. The absolute 
and relative errors for the {\tt rk8pd} routine in the 
simulations have been set to $10^{-8}$. 
The plot demonstrates that for those settings $d_k=0$
is satisfied by the numerical solutions up to $\sim 3\times 10^{-5}$ at
the end of the integration range. Thereby the accuracy is better for
the lowest modes $k=1,...,10$ than for the modes $k=51,...,60$. This
is partly due to the fact that the higher modes are more affected by the 
truncation of the infinite system at $k_{\rm max}=60$ than the lowest
modes. The accuracy for the intermediate modes $k=11,...,50$ lies in between 
the two ``bands'' visible in the plot. The remaining 
Bogoliubov relations are satisfied with at least the same accuracy 
demonstrating that the numerical errors are small compared
to the values of the particle numbers itself. We consider $d_k$ as 
the determining measure for the accuracy of 
the numerical calculations which can be easily enhanced further by 
increasing the preset accuracy of the integration routine. 

The numerical values for $N(t=249.5)$ and $N(t=250.0)$  
summarized in the table in Fig. \ref{figure 4} are shown with 
seven decimal places. Varying the cut-off between $k_{\rm max}=30$ 
and $60$ both values change only in the last two of the shown 
seven decimal places and therefore the variation in $N(t)$ when changing $k_{\rm max}$ 
is smaller than $10^{-5}$, i.e. smaller than the numerical 
error in the Bogoliubov relations. This demonstrates that the 
convergence of the numerical values for $N(t)$ is sufficiently good.  
Furthermore, because $\dot{l}(t=250)=3\pi\epsilon$ ($l_0=1$) 
we can conclude that spurious effects caused by discontinuities
in the velocity (matching problem) are indeed negligibly small  
for the parameters considered ($\epsilon\ll 1$).

\end{appendix}



\end{document}